\RequirePackage{ifpdf}
\newif\ifpdf
\ifx\pdfoutput\undefined
\pdffalse
\else
\pdfoutput=1
\pdftrue
\fi

\ifpdf
\documentclass[pdftex]{llncs}
\else
\documentclass[dvips]{llncs}
\fi

\usepackage[T1]{fontenc}
\usepackage{ae, aecompl}
\usepackage{wrapfig}
\usepackage{xspace}
\usepackage{algorithm}
\usepackage{algorithmicx}
\usepackage{algpseudocode}
\usepackage{float}
\newfloat{algorithm}{t}{lop}
\usepackage[vflt]{floatflt}
\usepackage{lineno}
\usepackage[usenames,dvipsnames]{color}
\usepackage{colortbl}

\usepackage{enumitem}
\usepackage{numprint}
\ifpdf
\usepackage[pdftex]{graphicx}
\DeclareGraphicsExtensions{.pdf,.png,.mps}
\else
\usepackage[dvips]{graphicx}
\usepackage[dvips]{color}
\fi

\usepackage{amssymb}
\usepackage{amsmath}
\usepackage{bibspacing}
\usepackage{xspace}

\usepackage{fancyhdr}
\newcommand{\eg}{e.\,g.\xspace}

\newcommand{\ie}{i.\,e.\xspace}

\newtheorem{app-theorem}{Example}

\newtheorem{consDefinition}[example]{Definition}
\newtheorem{consTheorem}[example]{Theorem}
\newtheorem{consProposition}[example]{Proposition}

\newcommand{\degree}{\operatorname{deg}}

\newcommand{\cond}{\operatorname{cond}}
\newcommand{\Cond}{\operatorname{Cond}}

\newcommand{\excond}{\operatorname{ex\_cond}}

\newcommand{\exalg}{\operatorname{ex\_alg}}

\newcommand{\RR}{\mathbb{R}}

\newcommand{\bigO}{\mathcal{O}}

\DeclareMathOperator{\intraWeight}{\it intraWeight}
\DeclareMathOperator{\interWeight}{\it interWeight}
\DeclareMathOperator{\subtreeVol}{\it subtreeVol}

\newcommand{\etal}{\textit{et al.}\xspace}

\newcommand{\wrt}{w.\,r.\,t.\xspace}

\newcommand{\metis}{\textsc{METIS}\xspace}

\newcommand{\kahip}{\textsc{KaHIP}\xspace}

\newcommand{\csch}[1]{\textcolor{blue}{[CS: #1]}\xspace}
\newcommand{\hmey}[1]{\textcolor{red}{[HM: #1]}\xspace}
\newcommand{\rgla}[1]{\textcolor{ForestGreen}{[RG: #1]}\xspace}
\renewcommand{\csch}[1]{}
\renewcommand{\hmey}[1]{}
\renewcommand{\rgla}[1]{}

\numberwithin{equation}{section}
\numberwithin{example}{section}
\numberwithin{table}{section}

\newcount\shortyear\newcount\shorthour\newcount\shortminute
\shorthour=\time\divide\shorthour by 60\shortyear=\shorthour
\multiply\shortyear by 60\shortminute=\time\advance\shortminute by
-\shortyear
\shortyear=\year\advance\shortyear by -1900

\def\zeit{\number\shorthour:\ifnum\shortminute<10 0\number\shortminute
\else\number\shortminute\fi}

\def\rightmark{{\small \today, \zeit{}}}

 \def\rightmark{}    

\begin{document}
\date{}

%
%
\title{Tree-based Coarsening and Partitioning \\ of Complex Networks}

\author{Roland Glantz \and Henning Meyerhenke \and Christian Schulz}

\institute{
Karlsruhe Institute of Technology (KIT), Karlsruhe, Germany}
\maketitle
%
\vspace{-4mm}
\begin{abstract}
  Many applications produce massive complex networks whose
  analysis would benefit from parallel processing.  Parallel
  algorithms, in turn, often require a suitable network partition.  For solving optimization tasks such as
  graph partitioning on large networks, multilevel methods are
  preferred in practice. Yet, complex networks pose  challenges to
  established multilevel algorithms, in particular to their coarsening
  phase.  

One way to specify a (recursive) coarsening of a graph is to
  rate its edges and then contract the edges as prioritized by the
  rating.  In this paper we (i) define weights for the edges of a
  network that express the edges' importance for connectivity, (ii)
  compute a minimum weight spanning tree $T^m$ \wrt these weights, and
  (iii) rate the network edges based on the conductance values of
  $T^m$'s fundamental cuts. To this end,  we also (iv) develop the first optimal
  linear-time algorithm to compute the conductance values of
  \emph{all} fundamental cuts of a given spanning tree.

  We integrate the new edge rating into a leading multilevel graph
  partitioner and equip the latter with a new greedy postprocessing
  for optimizing the maximum communication volume (MCV).  Experiments
  on bipartitioning frequently used benchmark networks show that the
  postprocessing already reduces MCV by 11.3\%. Our new edge rating
  further reduces MCV by 10.3\% compared to the previously best rating
  with the postprocessing in place for both ratings. In total, with a
  modest increase in running time, our new approach reduces the MCV of
  complex network partitions by 20.4\%.
\end{abstract}
\vspace{-4mm}


%
%
\section{Introduction}
\label{sec:intro}
%
Complex networks such as social networks or web graphs have
become a focus of investigation recently~\cite{costa2011analyzing}.
Such networks are often scale-free, \ie they have a power-law
degree distribution with many low-degree vertices and few high-degree vertices.
They also have a small diameter (small-world property), 
so that the whole network is discovered within a few hops from any vertex.
Complex networks arise in a variety of applications; several of them generate massive data sets.
As an example, the social network Facebook currently contains a billion active 
users (\url{http://newsroom.fb.com/Key-Facts}).
On this scale many algorithmic tasks benefit from parallel processing.
The efficiency of parallel algorithms on huge networks, in turn,
is usually improved by \emph{graph partitioning} (GP).

Given a graph $G=(V,E)$ and a number of blocks $k>0$, the GP problem asks for a
division of $V$ into $k$ pairwise disjoint subsets $V_1, \dots, V_k$
(\emph{blocks}) such that no block is larger than
%
%
$(1+\varepsilon)\cdot \left\lceil\frac{|V|}{k}\right\rceil,$
%
where $\varepsilon \geq 0$ is the allowed imbalance. When GP is used for parallel processing, each 
processing element (PE) usually receives one block, and edges running between two blocks model communication
between PEs. The most widely used objective function (whose minimization is $\mathcal{NP}$-hard) 
is the \emph{edge cut}, the total weight of the edges between different blocks.
However, it has been pointed out more than a decade ago~\cite{Hendrickson_graphpartitioning} that 
the determining factor for modeling the communication cost of parallel iterative graph algorithms is the
\emph{maximum communication volume} (MCV), which has received growing attention 
recently, \eg in a GP challenge~\cite{BaderMSW12dimacs}. 
MCV considers the worst communication volume taken over all blocks $V_p$
($1 \leq p \leq k$) and thus penalizes imbalanced communication:
$MCV(V_1, \dots, V_k) := \max_p \sum_{v \in V_p} |\{ V_i ~|~ \exists \{u, v\} \in E \mbox{ with } u \in V_i  \neq V_p\}|.$
%
Note that parallel processing is only one of many applications for graph partitioning; more can
be found in recent surveys~\cite{GPOverviewBook,SPPGPOverviewPaper}.

All state-of-the-art tools for partitioning very large graphs in
practice rely on the multilevel approach~\cite{SPPGPOverviewPaper}.
In the first phase a hierarchy of graphs $G_{0},\dots,G_{l}$ is built
by recursive coarsening. $G_{l}$ is supposed to
 be very small in
size, but similar in structure to the input 
 $G_{0}$. In the second
phase a very good initial solution for $G_{l}$
 is computed. In the
final phase, the solution is prolongated to the next-finer
 graph,
where it is improved using a local improvement algorithm. This
process of prolongation and local improvement is repeated up to
$G_{0}$.

Partitioning static meshes and similar non-complex networks 
this way is fairly mature. Yet, the structure of complex networks (skewed degree
distribution, small-world property) distinguishes complex networks from traditional
inputs and makes finding small cuts challenging with current tools.

One reason for the difficulties of established multilevel graph
partitioners is the coarsening phase. Most tools rely on edge
contractions for coarsening. Traditionally, only edge weights have
guided the selection of the edges to be
contracted~\cite{karypis1998fast}. Holtgrewe
\etal~\cite{HoltgreweSS10engineering} recently presented a two-phase
approach that makes contraction more systematic by separating two
issues: An \emph{edge rating} and a \emph{matching algorithm}. The
rating of an edge indicates how much sense it makes to contract the
edge. The rating then forms the input to an approximate maximum weight
matching algorithm, and the edges of the resulting matching are
contracted.  As one contribution of this paper, we define a new edge
rating geared towards complex network partitions with low MCV.



%

\paragraph{Outline and Contribution.}
\label{subsec:contribution}
%
After the introduction we sketch the state of the art
(Section~\ref{sec:soa}) and settle necessary notation
(Section~\ref{sec:prelim}).
Our first technical contribution, described briefly in Section~\ref{sec:GP},
results from our goal to minimize MCV rather than the edge cut: We equip 
a leading multilevel graph partitioner with  greedy postprocessing that trades in small edge cuts for small MCVs.

%
Our main contributions follow in Sections~\ref{sec:BI} and~\ref{sec:effCond}.
The first one is a new edge rating, designed for complex networks by
combining local and non-local information. Its rationale is to find moderately
balanced cuts of high quality quickly (by means of the clustering measure
\emph{conductance}~\cite{Kannan04clustering} and its loose connection to MCV via isoperimetric graph partitioning~\cite{GradyS06isoperimetric}) and to use this information to indicate whether
an edge is part of a small cut or not. Finding such cuts is done by evaluating 
conductance for all \emph{fundamental cuts} of a 
minimum spanning tree of the input graph with carefully chosen edge weights. 
(a fundamental cut is induced by the removal of exactly one 
spanning tree edge, cf.\ Section~\ref{sec:prelim}).
%
The second main contribution facilitates an efficient computation of our new
edge rating. We present the first optimal linear-time algorithm to
compute the conductance values of all fundamental cuts of a spanning
tree. 

We have integrated both MCV postprocessing and our new edge rating
$\excond(\cdot)$ into \kahip~\cite{kaHIPHomePage,sandersschulz2013}, a
state-of-the-art graph partitioner with a reference implementation of
the edge rating $\exalg(\cdot)$, that yielded the best quality for
complex networks so far (see Section~\ref{sec:soa}).

Experiments in Section~\ref{sec:results} show that greedy MCV
postprocessing improves the partitions of our complex network
benchmark set in terms of MCV by 11.3\% on average with a comparable
running time.

Additional extensive bipartitioning experiments (MCV postprocessing
included) show that, compared to $\exalg(\cdot)$, the fastest variant
of our new edge rating further improves the MCVs by 10.3\%, at the
expense of an increase in running time by a factor of
1.79. Altogether, compared to previous work on partitioning complex
networks with state-of-the-art methods~\cite{Safro2012a}, the total
reduction of MCV by our new techniques amounts to 20.4\%.


%
%
%
\section{State of the Art}
\label{sec:soa}
Multilevel graph partitioners such as \metis~\cite{karypis1998fast} and \kahip~\cite{kaHIPHomePage,sandersschulz2013} (more are described in recent surveys~\cite{GPOverviewBook,SPPGPOverviewPaper}) typically employ recursive
coarsening by contracting edges, which are often computed as those of a matching.
Edge ratings are important in guiding the matching
algorithm; a successful edge rating is

\begin{align}
\vspace{-1ex}
expansion^{*2}(\{u, v\}) &= \omega(\{u, v\})^2 / (c(u) c(v))\label{eq:exp2},
\end{align}

\noindent where the weights of the vertices $u, v \in V$ and of the edges $\{u,
v\} \in E$ are given by $c(\cdot)$ and $\omega(\cdot)$, respectively~\cite{HoltgreweSS10engineering}.

To broaden the view of the myopic rating above (it does not look beyond its
incident vertices), Safro
\etal~\cite{Safro2012a} precompute the algebraic distance
$\rho_{\{u, v\}}$~\cite{Chen2011a} for the end vertices
of each edge $\{u, v\}$ and use the edge rating

\begin{align}
\label{eq:expalg}
\vspace{-1ex}
\exalg(\{u, v\}) = (1/\rho_{\{u,v \}}) \cdot expansion^{*2}(u, v)
\end{align}

\noindent For graphs with power-law degree distributions, $\exalg(\cdot)$ yields
considerably higher partition quality than $expansion^{*2}(\cdot)$~\cite{Safro2012a}.
This is due to the fact that algebraic distance expresses a semi-local connection
strength of an edge $\{u, v\}$~\cite{Chen2011a}. Specifically, $\rho_{\{u, v\}}$ is
computed from $R$ randomly initialized vectors that are smoothed by a Jacobi-style
over-relaxation for a few iterations. 
The idea is that the vector entries associated with
well-connected vertices even out more quickly than those of
poorly connected vertices. Thus, a high value of $\rho_{\{u, v\}}$
indicates that the edge $\{u, v\}$ constitutes a bottleneck and should
not be contracted. 

Another strategy for matching-based multilevel schemes in complex networks 
(\eg for agglomerative clustering~\cite{Fagginger:2013gc}) is to match unconnected
vertices at 2-hop distance in order to eliminate star-like structures.
Also, alternatives to matching-based coarsening exist, \eg
weighted aggregation schemes~\cite{ChevalierS09comparison,MeyerhenkeMS09graph}.



Pritchard and Thurimella~\cite{Pritchard2011a} use a spanning tree to
sample the {\em cycle space} of a graph in a uniform way and thus find
small cuts (consisting of a single edge, two edges or a cut vertex)
with high probability~\cite{Pritchard2011a}. Our method uses a minimum
weight spanning tree on a graph with carefully chosen edge weights. Moreover,
we sample the {\em cut-space}. The aim of the
sampling is to create a collection $\mathcal{C}$ of moderately balanced
cuts which form the basis of our new edge rating.


We integrate our new algorithms into \kahip~\cite{kaHIPHomePage,sandersschulz2013}.
\kahip focuses on solution quality and has been shown recently to be
a leading graph partitioner for a wide variety of graphs such as
road networks, meshes, and complex networks~\cite{dissSchulz}.
It implements several advanced multilevel graph partitioning
algorithms, meta-heuristics, and sophisticated local improvement schemes.

%
%
%
\section{Preliminaries}
\label{sec:prelim}
Let $G=(V,E, \omega)$ be a finite, undirected, connected, and simple
graph. Its edge weights are given by $\omega: E \mapsto \RR^+$. We
write $\omega_{u,v}$ for $\omega(\{u, v\})$ and extend $\omega$ to
subsets of $E$ through $\omega(E') = \sum_{e \in E'} \omega(e)$.

For subsets $V_1$, $V_2$ of $V$ with $V_1 \cap V_2 = \emptyset$, the
set $S(V_1, V_2)$ consists of those edges in $E$ that have one end
vertex in $V_1$ and the other end vertex in $V_2$. If, in addition to
$V_1 \cap V_2 = \emptyset$, it holds that (i) $V = V_1 \cup V_2$ and
(ii) $V_1, V_2 \not= \emptyset$, then the pair $(V_1,V_2)$ is called a
{\em cut} of $G$, and $S(V_1, V_2)$ is called the {\em cut-set} of
$(V_1,V_2)$. The weight of a cut $(V_1,V_2)$ is given by
$\omega(S(V_1, V_2))$. The \emph{volume} of any subset $V'$ of $V$ is
the total weight of the edges incident on $V'$ (which equals the sum over the
weighted degrees of the vertices in $V'$):

\begin{equation}
\label{eq:def:vol}
vol(V') = \omega(\{e=\{v', v\} \in E \mid v' \in V', v \in V\}),
\end{equation}

\begin{consDefinition}[Fundamental cut, cut-set $S_T(e_T)$, $\cond(e_T, T)$]~\\
\label{nota:funda_condu}
\noindent Let $T$ be a spanning tree of $G$, and let $e_T \in
E(T)$. If $T_1$ and $T_2$ are the connected components (trees) of the
graph $(V, E(T) \setminus \{e_T\})$, then $(V(T_1), V(T_2))$ is the
fundamental cut of $G$ with respect to $T$ and $e_T$, and

\begin{equation}
\label{eq:funda}
S_T(e_T) = S(V(T_1), V(T_2)).
\end{equation}

\noindent is the fundamental cut-set of $G$ with respect to $T$ and
$e_T$. Conductance is a common quality measure in graph
clustering~\cite{Kannan04clustering}. Its value for $(V(T_1), V(T_2))$ is

\begin{equation}
\label{eq:condu}
\cond(e_T, T) = \cond(V_1, V_2) = \frac{\omega(S_T(e_T))}{\min\{vol(V(T_1)), vol(V(T_2))\}}
\end{equation}

\end{consDefinition}

%
%
%
\section{Greedy MCV Optimization}
\label{sec:GP}
%
The ultimate applications we target with our graph partitioning algorithm are iterative parallel algorithms executed on complex
networks. As argued in Section~\ref{sec:intro}, the maximum communication volume (MCV)
is a more accurate optimization criterion than the edge cut.
%
The graph partitioner \kahip has so far solely focused on the edge
cut, though. That is why, as a new feature, we equip \kahip with a
postprocessing that greedily optimizes MCV. This postprocessing is
executed after local improvement on the finest level of the multilevel
hierarchy and works in rounds.
%
In each round, we iterate over all boundary vertices of the input
partition in a random order and check whether moving the vertex from its
own block to the opposite block reduces or keeps the MCV value.
If this is the case, the current vertex will be moved to the
opposite block.  One round of the algorithm can be implemented in
$\mathcal{O}({\vert E \vert})$ time (see Section~\ref{sec:mcv-apx} in the appendix
for more details).  The total number of rounds of the algorithm is a tuning parameter.  After
preliminary experiments we have set it to 20.

%
%
%
\section{A New Conductance-based Edge Rating for Partitioning}
\label{sec:BI}
%
%
An edge rating in a multilevel graph partitioner should yield a low
rating for an edge $e$ if $e$ is likely to be contained in the cut-set
of a ``good'' cut, \eg if the cut-set consists of a bridge.

In our approach a good cut is one that (i) has a low conductance and
(ii) is at least moderately balanced. In complex networks (i) does not
always imply (ii) (see below).  A loose connection between conductance
and MCV in bipartitions can be established via isoperimetric graph
partitioning~\cite{GradyS06isoperimetric}.  Our approach to define an
edge rating and use it for partitioning is as follows.

\begin{enumerate}
\item Generate a collection $\mathcal{C}$ of moderately balanced
  bipartitions (cuts of $G$) that contain cuts with a low conductance
  value.
\item Define a measure $\Cond(\cdot)$ such that $\Cond(e)$ is low
  [high] if $e$ is [not] contained in the cut-set of a cut in
  $\mathcal{C}$ with low conductance.
\item Instead of multiplying the edge rating $expansion^{*2}(\{u,
  v\})$ with the factor $(1/\rho_{\{u, v\}})$ as in~\cite{Safro2012a},
  we replace one of the two (identical) myopic factors $\omega(\{u,
  v\})$ in $expansion^{*2}(\{u, v\})$ by the more far-sighted factor
  $\Cond(\cdot)$. This yields the new edge rating

  \begin{equation}
  \label{eq:excond}
  \excond(\{u, v\}) = \omega(\{u, v\}) \Cond(\{u, v\}) / (c(u) c(v))   
  \end{equation}

  The higher $\Cond(e)$, the higher $\excond(e)$, and thus the higher
  the chances for $e$ to be contracted during coarsening.
\item Run a multilevel graph partitioner capable of handling edge ratings
such as \kahip with $\excond(\cdot)$.
\end{enumerate}

\noindent To specify $\excond(\cdot)$, we need to define $\mathcal{C}$
and $\Cond(\cdot)$.

\begin{figure}[tb]
\begin{centering}
\includegraphics[scale=0.12]{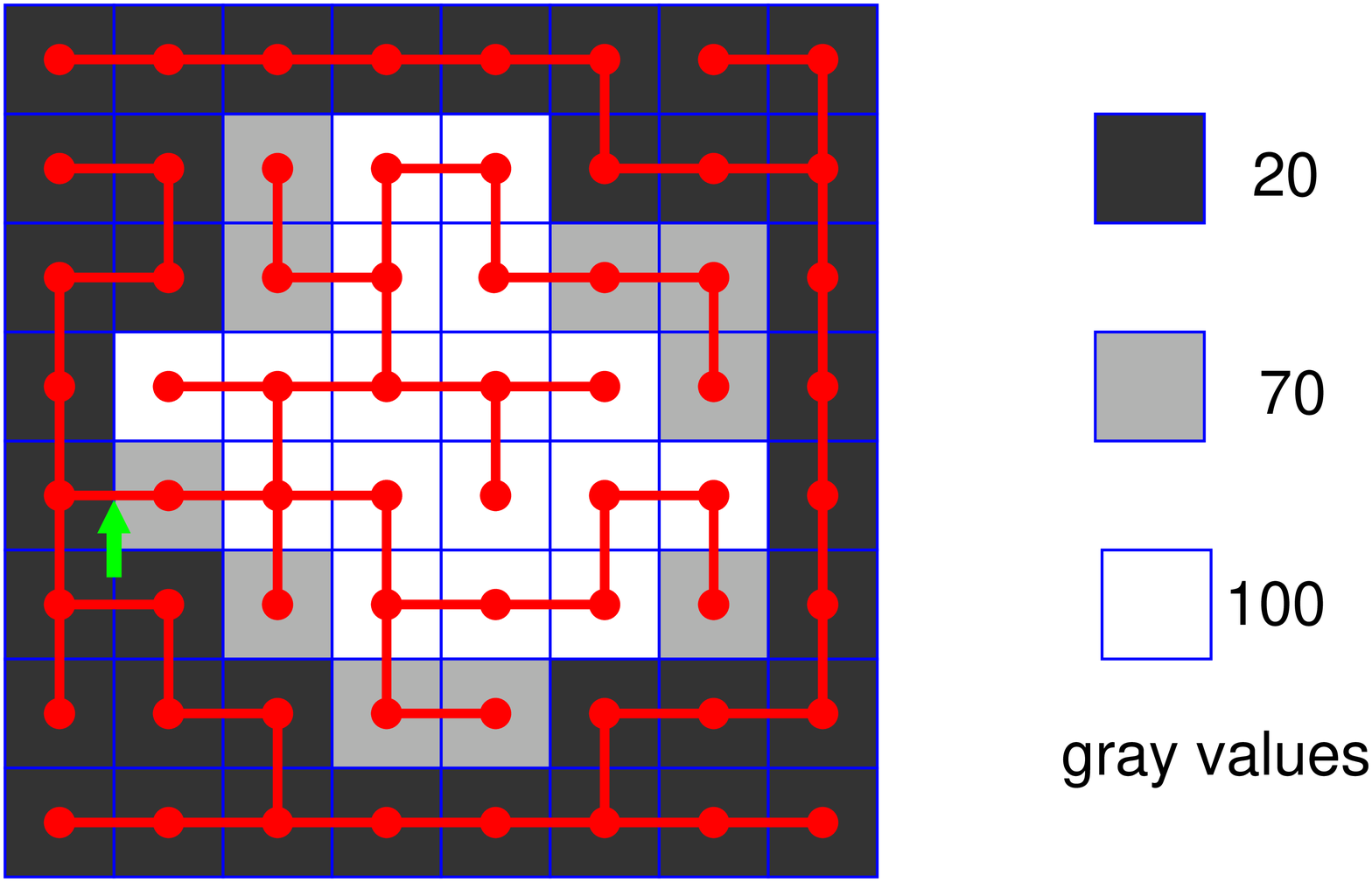}
\hspace{12ex}
\includegraphics[scale=0.20]{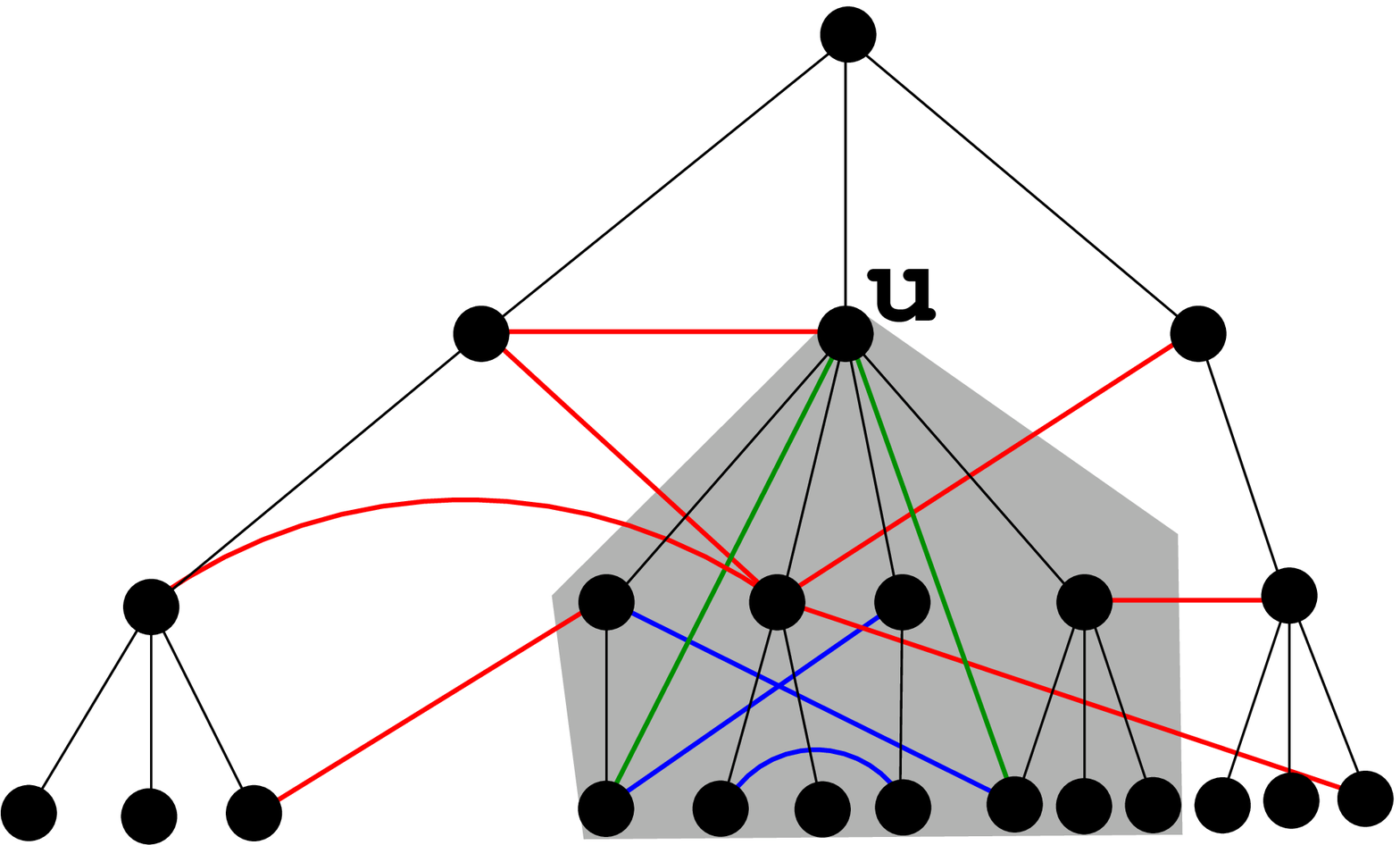}
\par\end{centering}
\caption{\label{fig:arrays} (a) Example of MST (red) in GBIS. For the
  green arrow see the text. (b) Vertex attributes $\intraWeight$ and
  $\interWeight$. Tree $T$ is formed by the black edges, and the
  subtree with root $u$ is contained in the shaded area. The weights
  of the blue and green edges contribute to $\intraWeight[u]$, and the
  weights of the red edges contribute to $\interWeight[u]$.}
\vspace{-0.5cm}
\end{figure}

\paragraph{Specifics of $\mathcal{C}$.}
For the definition of $\mathcal{C}$, we resort to a
basic concept of graph-based clustering, \ie the use of minimum weight
spanning trees (MSTs). We describe this concept in the context of
graph-based image segmentation
(GBIS)~\cite{Felzenszwalb2004a,Wassenberg2009a} for illustration
purposes (see Figure~\ref{fig:arrays}a).

%
%
In GBIS one represents an image by a graph $G$ whose vertices [edges]
represent pixels [neighborhood relations of pixels]. The edges are
equipped with weights that reflect the contrast between the gray
values at the edges' end vertices. An MST $T^m$ of $G$ with respect to
contrast has the following property
(see~\cite[Thm.~4.3.3]{Jungnickel05a}).  The contrast value associated
with any $e \in E(T^m)$ is minimal compared to the contrast values of
the edges in the fundamental cut-set $S_{T^m}(e)$ (see
Eq.~\ref{eq:funda}). Thus, for any $e \in E(T^m)$ with a high contrast
value (see the green arrow in Figure~\ref{fig:arrays}a), the
fundamental cut $S_{T^m}(e)$ yields a segmentation into two regions
with a high contrast anywhere on the common border.

Here, we arrive at a collection $\mathcal{C}$ of $\vert V \vert - 1$
moderately balanced bipartitions (cuts) of $G$ by (i) computing
connectivity-based contrast values for the edges of $G$, (ii)
computing an MST $T^m$ of $G$ \wrt these values, and (iii) letting
$\mathcal{C}$ consist of $G$'s fundamental cuts \wrt $T^m$. The
contrast value of an edge $e = \{u, v\}$ should be low [high] if the
edge is indispensable for ``many'' connections via shortest
paths. Thus, the higher the contrast, the stronger the negative effect
on $G$'s connectivity if $e$ is cut, and thus the more reasonable it
is to cut $e$.
To define the contrast values, we generate a random collection
$\mathcal{T}$ of breadth-first-traversal (BFT) trees. A tree in
$\mathcal{T}$ is formed by first choosing a root randomly. As usual,
we let the trees grow out of the roots using a queue, but we process
the edges incident on a vertex in a randomized order. Alternatively,
SSSP trees may be used if edge weights are to be included.

Let $n_{\mathcal{T}}(u, v)$ denote the number of trees in
$\mathcal{T}$ that contain $e$ and in which $u$ is closer to the
tree's root than $v$ ($u$ and $v$ cannot have the same distance to the
root). We set the {\em contrast} value of an edge $\{u, v\}$ to

\begin{equation}
\label{eq:gamma}
\gamma(\{u,v\}) = \min\{n_{\mathcal{T}}(u, v), n_{\mathcal{T}}(v, u)\}.
\end{equation}

Just considering the number of trees in $\mathcal{T}$ which contain
$e$, turned out to yield poorer partitions than using
Eq.~\ref{eq:gamma}. We believe that this is due to small subgraphs
which are connected to the graphs' ``main bodies'' via very few
edges. Just considering the number of trees in $\mathcal{T}$ which
contain such an edge would result in a high contrast of the edge
although the cut is far from moderately balanced. Even worse, the
conductance of the cut may be small (\eg if the cut-set contains only
one edge). This would protect edges in cut-sets of very unbalanced
cuts from being contracted --- an undesired feature.

\paragraph{Specifics of $\Cond(\cdot)$.}
Our plan is to define a measure $\Cond(\cdot)$ such that $\Cond(e)$ is
low [high] if $e$ is [not] contained in the cut-set of a cut in
$\mathcal{C}$ with low conductance. Hence, we set

\begin{equation}
\label{eq:Cond}
\Cond(e) = \min_{C \in \mathcal{C}, e \in S(C)}(\cond(C)), 
\end{equation}

\noindent where $S(C)$ denotes the cut-set of the cut $C$. Let $FC_e$
denote the set of edges in the (fundamental) cycle that arises if $e$
is inserted into $T^m$. Then, the cuts $C \in \mathcal{C}$ with $e \in
S(C)$ (see Eq.~\ref{eq:Cond}) are precisely the fundamental cuts
$S_T(e_T)$ (see Eq.~\ref{eq:funda}) with $e_T \in FC_e$ and $e_T \neq
e$. Note that $e$ is the only edge in $FC_e$ that is not in
$E(T^m)$. This suggests to first compute the $\Cond$-values for all
edges $e_T \in E(T^m)$ as specified in Section~\ref{sec:effCond}.  For
$e \notin E(T^m)$ the value of $\Cond(e)$ is then obtained by forming
the minimum of the $\Cond$-values of $FC_e \setminus \{e\}$. If $e =
\{u, v\}$, then $FC_e \setminus \{e\}$ is the set of edges on the
unique path in $T^m$ that connects $u$ to $v$.



%
%
%
\section{An $\bigO(\vert E \vert)$-Algorithm for Computing All
  $\cond(e_T, T)$ }
\label{sec:effCond}
In this section we demonstrate how, for a rooted given spanning tree
$T$ of a graph $G(V, E)$, one can compute all conductance values
$\cond(e_T, T)$, $e_T \in E(T)$, in time $\bigO(\vert E \vert)$ (the
root can be chosen randomly). This algorithm facilitates an efficient
computation of the edge rating introduced in the previous section.
The key to achieving optimal running time is to aggregate information
on fundamental cuts during a postorder traversal of $T$. The
aggregated information is kept in the three vertex attributes
$\subtreeVol$, $\intraWeight$ and $\interWeight$ defined in
Definition~\ref{def:attributes} below.
Technically, the three vertex attributes take the form
of arrays, where indices represent vertices.

\begin{consDefinition} 
\label{def:attributes}
Let $C_T(u)$ be the children of vertex $u$ in $T$.  Moreover, let
$T(u)$ denote the subtree rooted at $u$ (including $u$), and let $D(u)$
(\emph{descendants of $u$}) denote the set that contains the vertices
of $T(u)$, \ie $D(u) = V(T(u))$. We use the following three vertex
attributes to aggregate information that we need to compute the
conductance values:
\begin{itemize}[noitemsep,nolistsep]
  \item $\subtreeVol[u] = vol(D(u))$.
  \item $\intraWeight[u]$ equals twice the total weight of all edges $e
    = \{v, w\}$ with (i) $v, w \in D(u)$, (ii) $v, w \neq u$ and (iii)
    the lowest common ancestor of $v$ and $w$ in $T$ is $u$ ({\color{blue}blue}
    edges in Figure~\ref{fig:arrays}b) plus the total weight of all
    edges not in $T$ with one end vertex being $u$ and the other end
    vertex being contained in $D(u)$ ({\color{ForestGreen}green} edges in
    Figure~\ref{fig:arrays}b).
  \item $\interWeight[u]$ equals the total weight of all edges not
    in $T$ with exactly one end vertex in $D(u)$ ({\color{red}red} edges in
    Figure~\ref{fig:arrays}b).
\end{itemize}
\end{consDefinition}
%
%
If $u$ has a parent edge $e_T$, Eq.~\ref{eq:condu} takes the form

\begin{equation}
\label{eq:condu2}
\cond(e_T, T) = \frac{\\interWeight[u] + \omega(e_T)}{\min\{\subtreeVol[u], vol(V)-\subtreeVol[u]\}}
\end{equation}

When computing $\subtreeVol$, $\intraWeight$ and $\interWeight$, we
employ two vertex labellings (stored in arrays indexed by the
vertices): $label[u]$ indicates the preorder label of $u$ in $T$, and
$maxLabelDescendants[u]$ indicates the maximum of $label[t]$ over all
$t \in T(u)$. We also need lowest common ancestors (LCAs). Queries
\textsc{LCA}($T$, $u$, $v$), \ie the LCA of $u$ and $v$ on $T$,
require constant time after an $\bigO(n)$-time
preprocessing~\cite{Bender:2000:LPR:646388.690192}.

We start by initializing labels and vertex attributes to arrays of
length $\vert V \vert$ with all entries set to $0$ (for details see
Algorithm~\ref{algo:BTE} in Section~\ref{sec:pseudocode} in the
appendix). Then we compute the entries of $label$ and
$maxLabelDescendants$ in a single depth-first traversal of $T$ and
perform the preprocessing for \textsc{LCA}($\cdot, \cdot,
\cdot$). Finally, we call a standard postorder traversal in $T$
starting at the root of $T$. When visiting a vertex, either one of the
subroutines \textsc{Leaf}$(\cdot)$ (see Algorithm~\ref{algo:Leaf} in
Section~\ref{sec:pseudocode} of the appendix) or
\textsc{NonLeaf}$(\cdot)$ (see Algorithm~\ref{algo:NonLeaf}) is called
depending on the vertex type.

\begin{algorithm}[tbp]
\caption{Procedure \textsc{NonLeaf}$(T, u)$ called during postorder traversal of $T$}
\label{algo:NonLeaf}
\begin{algorithmic}[1]
\State $parentEdge \gets \textsc{undefined\_edge}$

\ForAll{$f = \{u, t\} \in E$}
  \If{$f \in E(T)$}
    \If{label[u] < label[t]}
      \State $\subtreeVol[u] \gets \subtreeVol[u] + \subtreeVol[t]$
      \State $\interWeight[u] \gets \interWeight[u] + \interWeight[t]$
    \Else
      \State $parentEdge \gets f$
    \EndIf
  \Else
  \If{$((label[t] < label[u])~\vee~(label[t] > maxLabelDescendants[u]))$}\\ 
    \Comment{equivalent to test if $t \notin D(u)$}
    \State $lca \gets LCA(T, u, t)$
    \State $\intraWeight[lca] \gets \intraWeight[lca] + \omega(f)$
    \State $\interWeight[u] \gets \interWeight[u] + \omega(f)$
  \EndIf
\EndIf
\EndFor
\State $\subtreeVol[u] \gets \subtreeVol[u] + vol(\{u\})$ \State
$\interWeight[u] \gets \interWeight[u] - \intraWeight[u]$;
\If{$parentEdge \not= \textsc{undefined\_edge}$} \State
$\cond(parentEdge, T) \gets \frac{\interWeight[u] +
  \omega(parentEdge)}{\min\{\subtreeVol[u], vol(V) - \subtreeVol[u]\}} $
\EndIf
%
\end{algorithmic}
\end{algorithm}

If $u$ is a leaf, Algorithm~\ref{algo:Leaf} sets $\subtreeVol[u]$ to
$vol(\{u\})$ and $\interWeight[u]$ to the total weight of all edges in
$E \setminus E(T)$ that are incident on $u$. Likewise, the entry
$\intraWeight[LCA(T, u,t)]$ is updated for any $t$ with $\{u, t\} \notin E(T)$. 

If $u$ is not a leaf (Algorithm~\ref{algo:NonLeaf}), and if $u$ has a parent edge in $T$, 
this edge is found in line 8 and the corresponding conductance value is computed in line 22 using $\subtreeVol[u]$ and
$\interWeight[u]$. The entry $\intraWeight[LCA(T, u,t)]$ is updated
multiple times until the postorder traversal ascends from $u$ towards
the root of $T$ (line 14). The
update of $interWeight$ is justified in the proof of
Theorem~\ref{prop:update} (see Section~\ref{sec:proofs-apx} 
(appendix), it also contains the proof of Proposition~\ref{prop:time}).
Eq.~\ref{eq:inter} in Theorem~\ref{prop:update}
guarantees that the conductance values computed in line 22 are correct. 





\begin{consTheorem}
\label{prop:update}
After having finished processing $u \in V$ in a traversal of $T$, the
equalities given below hold (where in the last one we assume that $u$
is not the root of $T$ and that $e_T$ is the parent edge of $u$ in
$T$).

\begin{align}
\subtreeVol[u] &= vol(D(u)),\label{eq:vol}\\
\intraWeight[u] &= {\color{blue}\sum_{c_i \neq c_j \in C(u)}\omega(S(D(c_i), D(c_j)))} \\
&~~~+ {\color{ForestGreen}\omega(S(D(u) \setminus \{u\}, \{u\}) \setminus E(T))}\mbox{~and}\label{eq:intra}\\
\interWeight[u] &= \omega(S_T(e_T)) - \omega(e_T)\label{eq:inter}.
\end{align}
\end{consTheorem}

\begin{consProposition}
\label{prop:time}
Given a rooted spanning tree $T$ of $G=(V, E)$, the computation
of all $\cond(e_T, T)$, $e_T \in E(T)$, takes $\bigO(|E|)$ time.
\end{consProposition}

%
%
%
\section{Experimental Results}
\label{sec:results}
%
\paragraph{Approach and settings.}

\begin{wraptable}[16]{r}{55mm}
\vspace{-8mm}
\scalebox{0.8}{
  \begin{tabular}{ l | r | r }
    Name & \#vertices & \#edges\\ \hline \hline
p2p-Gnutella          & \numprint{6405}   & \numprint{29215}\\\hline
PGPgiantcompo         & \numprint{10680}  & \numprint{24316}\\\hline
email-EuAll           & \numprint{16805}  & \numprint{60260}\\\hline
as-22july06           & \numprint{22963}  & \numprint{48436}\\\hline
soc-Slashdot0902      & \numprint{28550}  & \numprint{379445}\\\hline
loc-brightkite\_edges & \numprint{56739}  & \numprint{212945}\\\hline
loc-gowalla\_edges    & \numprint{196591} & \numprint{950327}\\\hline
coAuthorsCiteseer     & \numprint{227320} & \numprint{814134}\\\hline
wiki-Talk             & \numprint{232314} & \numprint{1458806}\\\hline
citationCiteseer      & \numprint{268495} & \numprint{1156647}\\\hline
coAuthorsDBLP         & \numprint{299067} & \numprint{977676}\\\hline
web-Google            & \numprint{356648} & \numprint{2093324}\\\hline
coPapersCiteseer      & \numprint{434102} & \numprint{16036720}\\\hline
coPapersDBLP          & \numprint{540486} & \numprint{15245729}\\\hline
as-skitter            & \numprint{554930} & \numprint{5797663}\\\hline
  \end{tabular}}
\caption{Complex networks used as benchmark set.}
\label{tab:graphs}
\end{wraptable}

The multilevel partitioner within the \kahip package has three
different algorithm configurations: strong, eco and fast. We use the
eco configuration since this configuration was chosen
in~\cite{Safro2012a}, too. The variable $\varepsilon$ in the balance
constraint 
is set to the common value $0.03$.

We evaluate the postprocessing and compare $\excond$ with $\exalg$ on
the basis of the 15 complex networks listed in Table~\ref{tab:graphs}
and further described in Table~\ref{tab:graphs_annot} 
(appendix). The networks are from two
popular archives~\cite{BaderMSW12dimacs,Leskovecxxxx}. The same
networks have been used previously in~\cite{Safro2012a} to evaluate
$\exalg$.

All computations are done on a workstation with two 8-core Intel(R)
Xeon(R) E5-2680 processors at 2.70GHz. Our code is implemented in
C/C++ and compiled with GCC 4.7.1.  Note that we do not exploit
parallelism here and run sequential experiments only.  First of all,
we focus in this paper on solution quality, not on speed. Secondly,
the standard of reference, $\exalg$, is also implemented sequentially.

Since the results produced by \kahip depend on many factors including random seeds,
we perform 50 runs with different seeds for each network and compute the following three {\em performance
  indicators}:

\begin{itemize}[noitemsep,nolistsep]
\item $minMCV$ and $avgMCV$: minimal and average MCV found by \kahip. 
\item $minCut$ and $avgCut$: minimal and average cut found by \kahip. 
\item $avgTime$: average time \kahip needs for the complete partitioning process.
\end{itemize}


\paragraph{Postprocessing results.}
For $\exalg$, the average reduction of avgMCV due to
postprocessing amounts to 11.3\% (see Table~\ref{tab:PPgainsMCV} in
Section~\ref{sec:exp-apx} of the appendix). Since the postprocessing
trades in small edge cuts for small MCVs, values for $minMCV$ and
$avgMCV$ [$minCut$ and $avgCut$] are with [without]
postprocessing. The increase in running time due to postprocessing is
negligible.
\vspace{-1.75ex}
\paragraph{Edge rating results.}
Intriguingly, using an asymptotically optimal Range Minimum Query
(RMQ) code (by Fischer and Heun~\cite{Fischer2006a}) within $\excond$
for the algorithms in Section~\ref{sec:effCond} 
does not decrease the running time. The straightforward asymptotically
slower algorithm is slightly faster (1.1\% in total) in our
experiments. To investigate this effect further, we compare the results on a set of 
non-complex networks, Walshaw's graph partitioning
archive~\cite{SoperWC04combined}. Again, the implementation of the (in
theory faster) RMQ algorithm does not play out, running time and
quality remain comparable. Therefore, the running times in all tables
refer to the implementation not using the Fischer/Heun RMQ code.

The edge rating $\excond$ depends on the number of random spanning
trees, \ie $\vert \mathcal{T} \vert$. To make this clear we write
$\excond_{\vert \mathcal{T} \vert}$ instead of $\excond$.

For a given network we measure the quality of the edge rating
$\excond_{\vert \mathcal{T} \vert}$ through (three) quotients of the
form (performance indicator using $\excond_{\vert \mathcal{T} \vert}$
divided by the same performance indicator using
$\exalg$). Tables~\ref{tab:MCV_20}, \ref{tab:MCV_100},
and~\ref{tab:MCV_200} in Section~\ref{sec:exp-apx} of the appendix
show the performance quotients of $\excond_{20}$, $\excond_{100}$ and
$\excond_{200}$. The geometric means of the performance quotients over
all networks are shown in Table~\ref{tab:meansQuot}.

\begin{table}[tb]
  \caption{Geometric means of the performance quotients minMCV, avgMCV
    and avgTime over all networks in
    Table~\ref{tab:graphs}. Number of trees: 20, 100 and 200. Reference is the edge
    rating $\exalg$. A quotient $< 1.0$ means that
    $\excond$ yields better results than $\exalg$.}
\begin{center}
  \begin{tabular}{ l | c c | c }
                               & minMCV            & avgMCV              & avgTime\\\hline \hline                                 
Ratios $\excond_{20}  / \exalg$ & \textbf{0.892}    & \textbf{0.897}      & \textbf{1.793}\\ \hline
Ratios $\excond_{100} / \exalg$ & \textbf{0.874}    & \textbf{0.893}      & \textbf{5.278}\\ \hline
Ratios $\excond_{200} / \exalg$ & \textbf{0.865}    & \textbf{0.890}      & \textbf{9.411}\\ \hline
\end{tabular}
\end{center}
\label{tab:meansQuot}
\vspace{-7mm} 
\end{table}

As the main result we state that buying quality through increasing
${\vert \mathcal{T} \vert}$ is expensive in terms of running time. The
rating $\excond_{20}$ already yields avgMCV that is 10.3\% lower than
avgMCV from $\exalg$ --- at the expense of a relative increase in
running time by only 1.79. The total reduction of average MCV from
postprocessing {\em and} replacing $\exalg$ by $\excond_{20}$ amounts
to 20.4\% (see Tables~\ref{tab:PPgainsMCV} and~\ref{tab:MCV_100} in
the appendix).

It is further interesting to note that, when we omit the
postprocessing step and compare the average edge cut instead of MCV,
$\exalg$ and $\excond$ perform comparably well. While $\excond$ yields
a slightly better minimum cut, $\exalg$ yields a slightly better
average cut (see Table~\ref{tab:cut_100} in Section~\ref{sec:exp-apx}
of the appendix).



%
%
%
\section{Conclusions and Future Work}
\label{sec:conclusions}
%
Motivated by the deficits of coarsening complex networks during
multilevel graph partitioning, we have devised a new edge rating for
guiding edge contractions. The new rating of an edge indicates whether
it is part of a good moderately balanced conductance-based cut or
not. To compute the necessary conductance values efficiently, we have
developed the first linear-time algorithm to compute the conductance
values of \emph{all} fundamental cuts of a spanning tree.
%
Our evaluation shows a significant improvement over a previously leading
code for partitioning complex networks. 
The new edge rating and additional greedy postprocessing \emph{combined}
result in a 20.4\% better maximum communication volume.

We would like to stress that good coarsening is not only of interest
for graph partitioning, but can be employed in many other methods and
applications that exploit hierarchical structure in networks. Future
work should investigate the concurrence of the contrast $\gamma$ and
the conductance values $\Cond $ --- possibly replacing $\gamma$ by an
even better contrast yet to be found. Our overall coarsening scheme is
agnostic to such a replacement and would require no further
changes. Moreover, we would like to extend our methods to an arbitrary
number of blocks. While the proposed edge rating should work out of
the box, the greedy MCV minimization has to be adapted to work
effectively for a larger number of blocks.


\paragraph*{Acknowledgments.}
We thank Johannes Fischer for providing an implementation of
the Range Minimum Query method presented in~\cite{Fischer2006a}.

\begin{footnotesize}
\bibliographystyle{abbrv}
\bibliography{roland,refs-parco}

\begin{thebibliography}{10}

\bibitem{BaderMSW12dimacs}
D.~A. Bader, H.~Meyerhenke, P.~Sanders, and D.~Wagner, editors.
\newblock {\em Graph Partitioning and Graph Clustering -- 10th DIMACS Impl.
  Challenge}, volume 588 of {\em Contemporary Mathematics}.
\newblock AMS, 2013.

\bibitem{Bender:2000:LPR:646388.690192}
M.~A. Bender and M.~Farach-Colton.
\newblock The {LCA} problem revisited.
\newblock In {\em Proc. of the 4th Latin American Symp. on Theoretical
  Informatics}, LATIN '00, pages 88--94, London, UK, UK, 2000. Springer-Verlag.

\bibitem{GPOverviewBook}
C.~Bichot and P.~Siarry, editors.
\newblock {\em Graph Partitioning}.
\newblock Wiley, 2011.

\bibitem{SPPGPOverviewPaper}
A.~Bulu\c{c}, H.~Meyerhenke, I.~Safro, P.~Sanders, and C.~Schulz.
\newblock {Recent Advances in Graph Partitioning}.
\newblock Technical Report ArXiv:1311.3144, 2014.

\bibitem{Chen2011a}
J.~{Chen} and I.~{Safro}.
\newblock Algebraic distance on graphs.
\newblock {\em SIAM J. Comput.}, 6:3468--3490, 2011.

\bibitem{ChevalierS09comparison}
C.~Chevalier and I.~Safro.
\newblock Comparison of coarsening schemes for multi-level graph partitioning.
\newblock In {\em Proc. Learning and Intelligent Optimization}, 2009.

\bibitem{costa2011analyzing}
L.~d.~F. Costa, O.~N. Oliveira~Jr, G.~Travieso, F.~A. Rodrigues, P.~R.
  Villas~Boas, L.~Antiqueira, M.~P. Viana, and L.~E. Correa~Rocha.
\newblock Analyzing and modeling real-world phenomena with complex networks: a
  survey of applications.
\newblock {\em Advances in Physics}, 60(3):329--412, 2011.

\bibitem{Fagginger:2013gc}
B.~O. {Fagginger Auer} and R.~H. Bisseling.
\newblock Graph coarsening and clustering on the {GPU}.
\newblock In {\em Graph Partitioning and Graph Clustering}. AMS and DIMACS,
  2013.

\bibitem{Felzenszwalb2004a}
P.~F. Felzenszwalb and D.~P. Huttenlocher.
\newblock Efficient graph-based image segmentation.
\newblock {\em Int. J. Comput. Vision}, 59(2):167--181, 2004.

\bibitem{Fischer2006a}
J.~Fischer and V.~Heun.
\newblock {Theoretical and Practical Improvements on the {RMQ}-Problem with
  Applications to {LCA} and {LCE}}.
\newblock In {\em Proc. 16th Symp. on Combinatorial Pattern Matching}, volume
  4009 of {\em LNCS}, pages 36--48. Springer, 2006.

\bibitem{GradyS06isoperimetric}
L.~Grady and E.~L. Schwartz.
\newblock Isoperimetric graph partitioning for image segmentation.
\newblock {\em IEEE Trans. Pattern Anal. Mach. Intell.}, 28(3):469--475, 2006.

\bibitem{Hendrickson_graphpartitioning}
B.~Hendrickson and T.~G. Kolda.
\newblock Graph partitioning models for parallel computing.
\newblock {\em Parallel Computing}, 26(12):1519--1534, 2000.

\bibitem{HoltgreweSS10engineering}
M.~Holtgrewe, P.~Sanders, and C.~Schulz.
\newblock Engineering a scalable high quality graph partitioner.
\newblock In {\em 24th Int. Parallel and Distributed Processing Symp. (IPDPS)},
  2010.

\bibitem{Jungnickel05a}
D.~{Jungnickel}.
\newblock {\em {Graphs, Networks and Algorithms}}, volume~5 of {\em {Algorithms
  and Computation in Mathematics}}.
\newblock Springer, Berlin, 2 edition, 2005.

\bibitem{Kannan04clustering}
R.~Kannan, S.~Vempala, and A.~Vetta.
\newblock On clusterings: Good, bad and spectral.
\newblock {\em J. of the ACM}, 51(3):497--515, 2004.

\bibitem{karypis1998fast}
G.~Karypis and V.~Kumar.
\newblock {A Fast and High Quality Multilevel Scheme for Partitioning Irregular
  Graphs}.
\newblock {\em SIAM J. on Scientific Computing}, 20(1):359--392, 1998.

\bibitem{Leskovecxxxx}
J.~{Leskovec}.
\newblock {S}tanford {N}etwork {A}nalysis {P}ackage ({SNAP}).

\bibitem{MeyerhenkeMS09graph}
H.~Meyerhenke, B.~Monien, and S.~Schamberger.
\newblock Graph partitioning and disturbed diffusion.
\newblock {\em Parallel Computing}, 35(10--11):544--569, 2009.

\bibitem{Pritchard2011a}
D.~{Pritchard} and R.~{Thurimella}.
\newblock Fast computation of small cuts via cycle space sampling.
\newblock {\em ACM Trans. Algorithms}, 7(4):46:1--46:30, 2011.

\bibitem{Safro2012a}
I.~{Safro}, P.~{Sanders}, and C.~{Schulz}.
\newblock Advanced coarsening schemes for graph partitioning.
\newblock In {\em Proc. 11th Int. Symp. on Experimental Algorithms}, pages
  369--380. Springer, 2012.

\bibitem{kaHIPHomePage}
P.~Sanders and C.~Schulz.
\newblock {KaHIP -- Karlsruhe High Qualtity Partitioning Homepage}.
\newblock {\url{http://algo2.iti.kit.edu/documents/kahip/index.html}}.

\bibitem{sandersschulz2013}
P.~Sanders and C.~Schulz.
\newblock {Think Locally, Act Globally: Highly Balanced Graph Partitioning}.
\newblock In {\em Proc. 12th Int. Symp. on Experimental Algorithms}, pages
  164--175. Springer, 2013.

\bibitem{dissSchulz}
C.~Schulz.
\newblock {\em {Hiqh Quality Graph Partititioning}}.
\newblock PhD thesis, Karlsruhe Institute of Technology, 2013.

\bibitem{SoperWC04combined}
A.~J. Soper, C.~Walshaw, and M.~Cross.
\newblock A combined evolutionary search and multilevel optimisation approach
  to graph partitioning.
\newblock {\em Journal of Global Optimization}, 29(2):225--241, 2004.

\bibitem{Wassenberg2009a}
J.~Wassenberg, W.~Middelmann, and P.~Sanders.
\newblock An efficient parallel algorithm for graph-based image segmentation.
\newblock In {\em Proc. 13th Int. Conf. Computer Analysis of Images and
  Patterns}, pages 1003--1010, 2009.

\end{thebibliography}
\end{footnotesize}

\newpage
\appendix
\section{Pseudocode of Algorithms in Section~\ref{sec:effCond}}
\label{sec:pseudocode}
\begin{algorithm}[!h]
\caption{Given a spanning tree $T$ of $G=(V, E)$ with root $r$,
  compute $\cond(e_T, T)$ for all $e_T \in E(T)$}
\label{algo:BTE}
\begin{algorithmic}[1]
\State Set $label, maxLabelDescendants, \subtreeVol, \intraWeight, \interWeight$ to $\vec{0}$

\State Compute $label[u]$ and $maxLabelDescendants[u]$ for all $u \in V$.

\State Perform LCA preprocessing

\State Postorder($T$, $r$)
\end{algorithmic}
\end{algorithm}

\begin{algorithm}[h!]
\caption{Procedure \textsc{Leaf}$(T, u)$ called during postorder traversal of $T$}
\label{algo:Leaf}
\begin{algorithmic}[1]
\State $\subtreeVol[u] \gets vol({\{u\}})$
\State $parentEdge \gets \textsc{undefined\_edge}$
\ForAll{$f = \{u, t\} \in E$}
  \If{$f \in E(T)$}
    \State $parentEdge \gets f$
  \Else
    \State $lca \gets $\textsc{LCA}$(T, u, t)$
    \State $\intraWeight[lca] \gets \intraWeight[lca] + \omega(f)$
    \State $\interWeight[u] \gets \interWeight[u] + \omega(f)$
  \EndIf
\EndFor
\If{$parentEdge \not= \textsc{undefined\_edge}$} \State
$\cond(parentEdge, T) \gets \frac{\interWeight[u] +
  \omega(parentEdge)}{\min\{\subtreeVol[u], vol(V) - \subtreeVol[u]\}} $
\EndIf
\end{algorithmic}
\end{algorithm}

\section{Proofs of Section~\ref{sec:effCond}}
\label{sec:proofs-apx}
\subsection{Proof of Theorem~\ref{prop:update}}
\begin{proof}
  We prove Eq.s~\ref{eq:vol},~\ref{eq:intra} and~\ref{eq:inter} one
  after the other. Colors correspond to the edge types introduced in
  Definition~\ref{def:attributes}.
\begin{enumerate}
\item Due to line 1 of Algorithm~\ref{algo:Leaf}, $\subtreeVol[u] =
  vol(\{u\}) = vol(D(u))$ for any leaf $u$ of $T$. If $u$ is not a
  leaf of $T$, we proceed by induction. Specifically, we assume inductively
  $\subtreeVol[c_i] = vol(D(c_i))$ for all children $c_i$ of $u$. Due to
  lines 5 and 19 of Algorithm~\ref{algo:NonLeaf}, $\subtreeVol[u] = \sum_{c_i \in C(u)}
  vol(D(c_i)) + vol(\{u\}) = vol(D(u))$.
\item Due to line 8 of Algorithm~\ref{algo:Leaf} and line 14 of
  Algorithm~\ref{algo:NonLeaf},

\begin{align*}
\intraWeight[u] &= \sum(\omega_{v,t} \mid v \in D(c_i) \mbox{~for some
  $c_i \in C(u)$},~\\
&~~~~~~~~~~~~~~~~~\{v, t\} \in E \setminus E(T),~LCA(v, t) = u)\\
&= \textcolor{blue}{\sum(\omega_{v,t} \mid v \in D(c_i) \mbox{~for
    some $c_i \in C(u)$},~t \in D(c_j))}\\
&~~~~~~~~~~~~~~~~\textcolor{blue}{\mbox{~for some $c_j \in C(u), c_j \neq c_i$}}~+\\
&~~~~\textcolor{ForestGreen}{\sum(\omega_{v,u} \mid v \in D(c_i) \mbox{~for
    some $c_i \in C(u)$})}~\\
&~~~~~~~~~~~~~~~~~\textcolor{ForestGreen}{\{v, u\} \in E \setminus E(T))}\\
&= \sum_{c_i \neq c_j \in C(u)}\omega(S(D(c_i), D(c_j))) +\\
&~~~~~~~~~~~~~~~~~\omega((E \setminus E(T)) \cap S(D(u) \setminus \{u\}, \{u\})).
\end{align*}

Note that a blue edge contributes twice to $\intraWeight[u]$ since it
is encountered from both endpoints. A green edge, on the other hand,
contributes only once.
\item Due to line 9 of Algorithm~\ref{algo:Leaf}, $\interWeight[u] =
  \sum_{e = \{u, t\}, e \in E \setminus E(T)}\omega(e) =
  \omega(S_T(e_T)) - \omega(e_T)$ for any leaf $u$ of $T$. If $u$ is
  not a leaf of $T$, let $e_i = \{u, c_i\}$ be the edges between $u$
  and its children $c_i$. In particular, $e_i \in E(T)$. By induction
  we may assume $\interWeight[c_i] = \omega(S_T(e_i)) -
  \omega(e_i)$. Thus, line 6 of Algorithm~\ref{algo:NonLeaf}, where we
  take the sum over all $\interWeight$ values of $u$'s children $c_i$,
  yields

\begin{align*}
\interWeight[u] &= \sum(\omega_{v, t} \mid \{v, t\} \in E \setminus
E(T) \wedge \exists c_i \in C(u): ~\\
&~~~~~~~~~~~~~~~~~~v \in D(c_i), t \notin D(c_i))\\
&= \textcolor{blue}{\sum(\omega_{v, t} \mid v \in D(c_i) \mbox{~for
    some $c_i \in C(u)$} \wedge t \in D(c_j))}\\
&~~~~~~~~~~~~~~~~\textcolor{blue}{\mbox{~for some $c_j \in C(u), c_j \neq c_i$})}\\
 &+ \textcolor{ForestGreen}{\sum(\omega_{v, u} \mid \{v, u\} \in E \setminus
  E(T) \wedge v \in D(u) \setminus \{u\})}\\
 &+ \textcolor{red}{\sum(\omega_{v, t} \mid \{v, t\} \in E \setminus
  E(T) \wedge v \in D(u) \setminus \{u\},~t \notin D(u))}
\end{align*}

Note that this is an intermediate result of $\interWeight[u]$. The blue
and green terms make up $\intraWeight[u]$, which we still have to
subtract. Moreover, the red term does not yet contain the (weights of)
edges in $T$ with one end vertex being $u$ and the other one not being
contained in $D(u)$. In the following, we replace the blue and the
green term by $\intraWeight[u]$ and rewrite the red term.

\begin{align*}
\interWeight[u] &= \intraWeight[u] + \textcolor{red}{\sum(\omega_{v, t}
  \mid \{v, t\} \in E \setminus E(T))},~\\
&~~~~~~~~~~~~~~~~~~~~~~~~~~~~~~~~~~~~~~~~~~\textcolor{red}{v \in D(u) \setminus \{u\},~t \notin D(u)}\\
&= \intraWeight[u] + \sum(\omega_{v, t} \mid \{v, t\} \in E \setminus
E(T),~\\
&~~~~~~~~~~~~~~~~~~~~~~~~~~~~~~~~~~~~~~~~~~v \in D(u),~t \notin D(u))~-\\
&~~~~\sum(\omega_{u, t} \mid \{u, t\} \in E \setminus E(T),~t \notin D(u))\\
&= \intraWeight[u] + \omega(S_T(e_T)) - \omega(e_T) -\\
&~~~~\sum(\omega_{u, t} \mid \{u, t\} \in E \setminus E(T),~t \notin D(u))
\end{align*}

Finally, line 15 of Algorithm~\ref{algo:NonLeaf} results in
$\interWeight[u] = \intraWeight[u] + \omega(S_T(e_T)) - \omega(e_T)$,
and line 20 of Algorithm~\ref{algo:NonLeaf} yields
Eq.~\ref{eq:inter}.
\end{enumerate}

\end{proof}

\subsection{Proof of Proposition~\ref{prop:time}}
\begin{proof}
All initialization and preprocessing steps can be done in $\bigO(n)$
time. During the postorder traversal of $T$ each $v \in V$ explores
its direct neighborhood, either in \textsc{Leaf} or in
\textsc{NonLeaf}. Two observations are crucial now. First, all
elementary operations within \textsc{Leaf} and \textsc{NonLeaf} take
constant time, including the LCA queries. Second, for each edge of
$G$ the respective operations are executed at most twice.
\end{proof}

\section{MCV Postprocessing}
\label{sec:mcv-apx}
We now show how one round of the postprocessing algorithm that
optimizes MCV can be implemented in $\mathcal{O}({\vert E \vert})$
time.  The crucial step is to decide if moving a vertex $v$ to the
opposite block reduces MCV in $O(\deg(v))$ time. To do so, we need a
few notations.  An \emph{internal vertex} is a vertex of the graph
which is not a boundary vertex, and the external degree of a vertex is
defined as the number of neighbors in the opposite block.  Let $(V_1,
V_2)$ be a bipartition of $G$ and, without loss of generality, let $v$
be a random boundary vertex from block $V_1$. During the course of the
algorithm, we keep track of the communication volumes of the
blocks. Let $C_1$ and $C_2$ be the initial communication volume of
$V_1$ and $V_2$, respectively. We do the following to decide if moving
$v$ to the opposite block reduces MCV.  First, we move $v$ to the
opposite block $V_2$. Afterwards, the communication volume $C_2$ is
reduced by the number of boundary vertices in $V_2$ that are also
neighbors of $v$ and become internal vertices after the movement.
Moreover, the communication volume $C_1$ is increased by the amount of
internal vertices in $V_1$ that become boundary vertices after $v$ is
moved to $V_2$.  Additionally, since we move $v$, the communication
volume $C_1$ is reduced by one, and if the number of neighbors of $v$
in $V_1$ is not zero, then $C_2$ is increased by one.  Note that we
can check in constant time if a vertex is a boundary vertex or an
internal vertex by storing the external degree of all vertices in an
array and updating the external degree of a vertex and its neighbors
when the vertex is moved.  We move $v$ back to its origin if the
movement did not yield an improvement in MCV.


\section{Test Set of Complex Networks}
\label{sec:test-set}
\begin{table}[h!]
\caption{Complex networks used for comparing $\excond$ and $\exalg$.}
\begin{center}
\scalebox{0.8}{
  \begin{tabular}{ l | r | r | r }
    Name & \#vertices & \#edges & Network Type\\ \hline \hline
p2p-Gnutella          & \numprint{6405}   & \numprint{29215}    & filesharing network\\\hline
PGPgiantcompo         & \numprint{10680}  & \numprint{24316}    & largest connected component in network of PGP users\\\hline
email-EuAll           & \numprint{16805}  & \numprint{60260}    & network of connections via email\\\hline
as-22july06           & \numprint{22963}  & \numprint{48436}    & network of autonomous systems in the internet\\\hline
soc-Slashdot0902      & \numprint{28550}  & \numprint{379445}   & news network\\\hline
loc-brightkite\_edges & \numprint{56739}  & \numprint{212945}   & location-based friendship network\\\hline
loc-gowalla\_edges    & \numprint{196591} & \numprint{950327}   & location-based friendship network\\\hline
coAuthorsCiteseer     & \numprint{227320} & \numprint{814134}   & citation network\\\hline
wiki-Talk             & \numprint{232314} & \numprint{1458806}  & network of user interactions through edits\\\hline
citationCiteseer      & \numprint{268495} & \numprint{1156647}  & citation network\\\hline
coAuthorsDBLP         & \numprint{299067} & \numprint{977676}   & citation network\\\hline
web-Google            & \numprint{356648} & \numprint{2093324}  & hyperlink network of web pages\\\hline
coPapersCiteseer      & \numprint{434102} & \numprint{16036720} & citation network\\\hline
coPapersDBLP          & \numprint{540486} & \numprint{15245729} & citation network\\\hline
as-skitter            & \numprint{554930} & \numprint{5797663}  & network of internet service providers\\\hline
  \end{tabular}}
\end{center}
\label{tab:graphs_annot}
\end{table}

\section{Details of Experimental Results}
\label{sec:exp-apx}

\begin{table}[tb]
  \caption{Effect of postprocessing on performance of $\exalg$. Numbers are ratios of
    the performance indicators minMCV and avgMCV with and without
    postprocessing. A ratio $r < 1.0$ means that postprocessing
    reduces MCV by $100 (1-r) \%$.}
\begin{center}
  \begin{tabular}{c | c  c}
                             & minMCV              &  avgMCV               \\\hline \hline
PGPgiantcompo                &  0.951              &   0.924               \\ \hline
as-22july06                  &  0.913              &   0.824               \\ \hline
email-EuAll                  &  0.858              &   0.859               \\ \hline
loc-brightkite\_edges        &  0.765              &   0.753               \\ \hline
p2p-Gnutella04               &  0.866              &   0.862               \\ \hline
soc-Slashdot0902             &  0.921              &   0.912               \\ \hline
citationCiteseer             &  0.939              &   0.919               \\ \hline
coAuthorsCiteseer            &  0.975              &   0.955               \\ \hline 
coAuthorsDBLP                &  0.905              &   0.898               \\ \hline
loc-gowalla\_edges           &  0.887              &   0.870               \\ \hline
web-Google                   &  0.965              &   0.926               \\ \hline
wiki-Talk                    &  0.994              &   0.974               \\ \hline
as-skitter                   &  0.939              &   0.937               \\ \hline
coPapersCiteseer             &  0.892              &   0.877               \\ \hline
coPapersDBLP                 &  0.856              &   0.841               \\ \hline  \hline
\textbf{Geometric mean}	     & \textbf{0.907}      & \textbf{0.887}        \\ \hline
\end{tabular}
\end{center}
\label{tab:PPgainsMCV}
\end{table}

\begin{table}[tb]
  \caption{Performance quotients of $\excond_{20}$ with postprocessing for minMCV, avgMCV
    and avgTime. Reference is $\exalg$ with postprocessing. A quotient $< 1.0$ means that
    $\excond_{20}$ yields better results than $\exalg$.}
\begin{center}
  \begin{tabular}{ c | c  c | c }
                         & minMCV  & avgMCV & avgTime\\
                        \hline \hline
PGPgiantcompo           &  0.962              &  1.004               &  3.068              \\ \hline
as-22july06             &  0.898              &  0.766               &  1.387              \\ \hline
email-EuAll             &  0.904              &  0.918               &  1.537              \\ \hline
loc-brightkite\_edges   &  0.714              &  0.714               &  1.190              \\ \hline
p2p-Gnutella04          &  1.003              &  1.000               &  2.255              \\ \hline
soc-Slashdot0902        &  0.991              &  0.999               &  3.045              \\ \hline
citationCiteseer        &  1.001              &  0.974               &  1.938              \\ \hline
coAuthorsCiteseer       &  1.090              &  1.053               &  2.842              \\ \hline 
coAuthorsDBLP           &  0.743              &  0.774               &  1.461              \\ \hline
loc-gowalla\_edges      &  0.608              &  0.607               &  0.642              \\ \hline
web-Google              &  0.820              &  0.729               &  3.479              \\ \hline
wiki-Talk               &  0.992              &  1.023               &  1.142              \\ \hline
as-skitter              &  0.769              &  0.760               &  0.924              \\ \hline
coPapersCiteseer        &  1.011              &  1.260               &  2.575              \\ \hline
coPapersDBLP            &  1.035              &  1.135               &  2.448              \\ \hline  \hline
\textbf{Geometric mean} &  \textbf{0.892}     &  \textbf{0.897}      &\textbf{1.793}       \\ \hline
\end{tabular}
\end{center}
\label{tab:MCV_20}
\end{table}

\begin{table}[tb]
  \caption{Performance quotients of $\excond_{100}$ with postprocessing for minMCV, avgMCV
    and avgTime. Reference is $\exalg$ with postprocessing. A quotient $< 1.0$ means that
    $\excond_{100}$ yields better results than $\exalg$.}
\begin{center}
  \begin{tabular}{ c | c  c | c }
                         & minMCV  & avgMCV & avgTime\\
                        \hline \hline
PGPgiantcompo           &  0.986              &  0.998               &  12.379              \\ \hline
as-22july06             &  0.750              &  0.760               &   2.373              \\ \hline
email-EuAll             &  0.874              &  0.917               &   3.183              \\ \hline
loc-brightkite\_edges   &  0.715              &  0.715               &   3.836              \\ \hline
p2p-Gnutella04          &  0.995              &  0.995               &   6.670              \\ \hline
soc-Slashdot0902        &  0.920              &  0.987               &  10.801              \\ \hline
citationCiteseer        &  1.017              &  0.972               &   7.485              \\ \hline
coAuthorsCiteseer       &  1.068              &  1.035               &  10.567              \\ \hline 
coAuthorsDBLP           &  0.737              &  0.776               &   5.282              \\ \hline
loc-gowalla\_edges      &  0.624              &  0.609               &   1.929              \\ \hline
web-Google              &  0.837              &  0.729               &  13.507              \\ \hline
wiki-Talk               &  0.987              &  1.015               &   1.194              \\ \hline
as-skitter              &  0.793              &  0.769               &   2.986              \\ \hline
coPapersCiteseer        &  0.991              &  1.245               &   8.567              \\ \hline
coPapersDBLP            &  0.970              &  1.118               &   7.951              \\ \hline  \hline
\textbf{Geometric mean} &  \textbf{0.874}     &  \textbf{0.893}      &\textbf{5.278}        \\ \hline
\end{tabular}
\end{center}
\label{tab:MCV_100}
\end{table}

\begin{table}[tb]
  \caption{Performance quotients of $\excond_{200}$ with postprocessing for minMCV, avgMCV
    and avgTime. Reference is $\exalg$ with postprocessing. A quotient $< 1.0$ means that
    $\excond_{200}$ yields better results than $\exalg$.}
\begin{center}
  \begin{tabular}{ c | c  c | c }
                         & minMCV  & avgMCV & avgTime\\
                        \hline \hline
PGPgiantcompo           &  0.976              &   1.000              &  24.258               \\ \hline
as-22july06             &  0.694              &   0.735              &   3.645               \\ \hline
email-EuAll             &  0.874              &   0.927              &   5.262               \\ \hline
loc-brightkite\_edges   &  0.692              &   0.711              &   7.215               \\ \hline
p2p-Gnutella04          &  0.997              &   0.997              &  12.313               \\ \hline
soc-Slashdot0902        &  0.913              &   0.959              &  20.883               \\ \hline
citationCiteseer        &  1.013              &   0.974              &  14.589               \\ \hline
coAuthorsCiteseer       &  1.026              &   1.035              &  20.412               \\ \hline 
coAuthorsDBLP           &  0.740              &   0.773              &  10.157               \\ \hline
loc-gowalla\_edges      &  0.615              &   0.611              &   3.574               \\ \hline
web-Google              &  0.843              &   0.729              &  26.421               \\ \hline
wiki-Talk               &  0.986              &   1.017              &   1.217               \\ \hline
as-skitter              &  0.775              &   0.760              &   5.588               \\ \hline
coPapersCiteseer        &  1.035              &   1.255              &  16.131               \\ \hline
coPapersDBLP            &  0.958              &   1.122              &  14.881               \\ \hline  \hline
\textbf{Geometric mean} &  \textbf{0.865}     &  \textbf{0.890}      &\textbf{9.411}        \\ \hline
\end{tabular}
\end{center}
\label{tab:MCV_200}
\end{table}

\begin{table}[tb]
  \caption{Performance quotients of $\excond_{100}$ without postprocessing for minCut, avgCut
    and avgTime. Reference is $\exalg$ without postprocessing. A quotient $< 1.0$ means that
    $\excond_{100}$ yields better results than $\exalg$.}
\begin{center}
  \begin{tabular}{ c | c  c | c }
                         & minMCV  & avgMCV & avgTime\\
                        \hline \hline
PGPgiantcompo           &  1.036              &  1.021               &  13.405               \\ \hline
as-22july06             &  0.858              &  0.904               &   2.387               \\ \hline
email-EuAll             &  0.976              &  0.894               &   3.248               \\ \hline
loc-brightkite\_edges   &  1.026              &  1.032               &   3.964               \\ \hline
p2p-Gnutella04          &  0.951              &  0.959               &   7.554               \\ \hline
soc-Slashdot0902        &  0.943              &  1.258               &  13.083               \\ \hline
citationCiteseer        &  0.986              &  0.941               &   8.000               \\ \hline
coAuthorsCiteseer       &  1.118              &  1.110               &  11.420               \\ \hline 
coAuthorsDBLP           &  0.795              &  0.885               &   5.554               \\ \hline
loc-gowalla\_edges      &  1.100              &  1.093               &   1.948               \\ \hline
web-Google              &  0.826              &  0.699               &  15.660               \\ \hline
wiki-Talk               &  1.032              &  1.033               &   1.194               \\ \hline
as-skitter              &  1.007              &  1.007               &   3.085               \\ \hline
coPapersCiteseer        &  1.150              &  1.391               &  13.578               \\ \hline
coPapersDBLP            &  1.034              &  1.252               &  12.045               \\ \hline  \hline
\textbf{Geometric mean} &  \textbf{0.984}     &  \textbf{1.018}      &\textbf{5.915}         \\ \hline
\end{tabular}
\end{center}
\label{tab:cut_100}
\end{table}

\end{document}